\documentclass[12pt]{article}
\usepackage{psfig}

\def\farcs{\hbox{$.\!\!^{\prime\prime}$}}
\def\reference{\parskip 0pt\par\noindent\hangindent 0.5 truecm}

\newdimen\digitwidth
\setbox0=\hbox{\rm0}
\digitwidth=\wd0
\catcode`#=\active
\def#{\kern\digitwidth}
\begin{document}
%
%
\title{FLAIR-II spectroscopy of two DENIS J-band galaxy samples}

%


\author{Gary A. MAMON $^{1,2}$ 
 Quentin A. PARKER $^{3,4}$ \and
 Dominique PROUST $^{2}$
} 

\date{}
\maketitle

{\center
$^1$ Institut d'Astrophysique de Paris, 98 bis Bd Arago, F--75014 Paris,
FRANCE\\{\tt gam@iap.fr}\\ 
$^2$ DAEC, Observatoire de Paris, F--92195 Meudon, FRANCE\\
$^3$ Institute for Astronomy, Royal Observatory, Edinburgh, UNITED KINGDOM\\
$^4$ Anglo-Australian Observatory, Coonabarabran, NSW 2357, AUSTRALIA

}

\vspace{-10cm}{\normalsize\noindent\emph{Publ. Astr. Soc. Aust., in press}}
\vspace{9.5cm}

%
\begin{abstract}
As a pilot survey for the forthcoming {\sf 6dF} Galaxy Redshift Survey,
spectroscopy of galaxies 
selected in the 1.2 micron $J$ waveband with the 
{\sf DENIS} imaging survey was performed at the {\sf UKST} telescope using the 
{\sf FLAIR~II} multi-object spectroscope.
69 galaxy 
redshifts were obtained in a high galactic latitude field and an additional
12 redshifts in a low galactic latitude ($b=-17^\circ$), obscured  field.
This spectroscopic followup of NIR selected galaxies illustrates the
feasibility of obtaining redshifts with optical spectra on galaxies selected
at much longer wavelengths.
It validated a very preliminary algorithm for star/galaxy separation for high
galactic latitude {\sf
DENIS} objects, with 99\% reliability for $J < 13.9$.
The {\sf FLAIR II} redshifts are in excellent agreement
with those, previously published, of 20 common galaxies.
However, the {\sf FLAIR II} redshift determinations presented
here required substantially
longer integration times to achieve 90\% completeness than expected
from previous optical surveys at comparable depth. This is 
mainly due to a 
degradation in overall
fibre throughput due to known problems with ageing of the
prism-cement-fibre interface with exposure to UV light.
In comparison to our high galactic latitude field, our low latitude 
(high extinction) field
required 2.5 times more exposure time for 
less than 50\% of successful redshift measurements.

Among the $J \leq 13.9$ galaxies with measured redshifts, only
$37\pm6\%$ display
emission lines, 
in comparison with 60\% of emission-line galaxies in optical samples of
comparable depth.
These galaxies  are, on average, half a magnitude bluer in $B-J$
than galaxies of the
same luminosity without emission lines.
We confirm a previous optically-based result that the fraction of galaxies
with emission lines increases rapidly with
decreasing galaxy luminosity.
The $J$-band luminosity function is estimated.
Our high latitude field displays a
concentration of galaxies at $cz \simeq 38\,000 \, \rm km \, s^{-1}$
suggesting a possible supercluster.
A radial velocity is reported for a galaxy lying near the projected centre of
the Abell 1434 cluster of galaxies, for which no cluster redshift is
currently available.

\end{abstract}


\bigskip

%
%

\section{Introduction}

The recent advances in detector technology in the near-infrared (hereafter
NIR)
domain have 
brought much enthusiasm to the field of
large-scale structure of the Universe.
Indeed, NIR light has the strong advantage of being much less
affected by extinction from dust grains, than is optical light, and as such,
near-infrared galaxy images show spiral galaxies in a much simpler light,
traced by their old stellar populations, with little dust obscuration in
spiral arms and in the galaxy centres (Zaritsky, Rix \& Rieke 1993;
H\'eraudeau, Simien \& Mamon 1996).
Moreover, NIR emission is less affected by recent bursts of star formation
within a galaxy, that 
can result in relatively small numbers of new hot OB stars
contributing a substantial fraction of the optical light of the
entire galaxy.
Finally, the transparency of dust grains to NIR photons means that our own
Galaxy will not obscure the view of external galaxies imaged in the NIR. 
However, within a
few degrees from the Galactic Plane, confusion from stars sets in and
prevents reliable star/galaxy separation (e.g. Jarrett et al. 2000b).

For these reasons, tracing the distribution of mass in the Universe with
galaxies becomes much more meaningful if the galaxies are selected in the NIR
domain.
Two very large NIR digital 
imaging surveys, {\sf DENIS} (e.g. Epchtein et al. 1999)
and {\sf 2MASS} (e.g. Jarrett et al. 2000a), are producing galaxy catalogues
that are essential for such cosmological studies in two dimensions.
Moreover, there is much science to be gained by using redshifts 
to give a third dimension. With this in mind, we have approval 
to use the soon to be commissioned {\sf 6dF} 
robotic multi-fibre spectroscope (Parker, Watson \& Miziarski 1998; Watson et
al. 2000)
on 
the {\sf UKST} to undertake the massive {\sf 6dF Galaxy Redshift Survey} 
(hereafter 
{\sf 6dFGRS}, e.g. Mamon 1998, 1999; see also 
{\tt http://msowww.anu.edu.au/\~{}colless/6dF})
of roughly $120\,000$ galaxies selected from {\sf
2MASS} and {\sf DENIS}.
This large survey is scheduled to begin in the southern autumn of
2001.

In a pilot study of {\sf 6dFGRS},
we tested the feasibility of measuring redshifts with optical spectroscopy of
NIR-selected 
galaxy samples.
For this, we observed two Schmidt-plate sized fields of galaxies, 
selected in the $J$
(1.25$\,\mu$m) band with early {\sf DENIS} observations.
The spectroscopy was performed at the {\sf UKST}
with 
the {\sf FLAIR II} multi-fibre instrument (e.g. Parker \& Watson 1995),
which has now been decommissioned pending the introduction of the automated
{\sf 6dF}. 
The results of this pilot study 
are reported in the present paper.

\section{Sample selection}

Galaxies were selected from preliminary {\sf DENIS} extractions performed
during 
February 1998, with a galaxy pipeline, based upon {\sf SExtractor} (Bertin \&
Arnouts 1996) and a simple preliminary
star/galaxy separation algorithm based upon a threshold of the ratio of peak
intensity to isophotal area ($I/A$), which, to 
first order, does not depend
on magnitude (Mamon et al. 1998).
This star/galaxy separation is performed in the {\sf DENIS}-$I$ band, which
is much more sensitive and has better spatial resolution 
than the {\sf
DENIS}-$J$ or $K$ 
bands.
The target galaxy samples were taken from the contiguous {\sf DENIS} strips
that overlap regions of {\sf UKST} standard survey fields 
F787 (12h00, $-5^\circ$) at high galactic latitude ($b = 55^\circ$), and
F20 (12h00, $-80^\circ$) at low galactic latitude ($b = -17^\circ$).

All objects in {\sf DENIS} \emph{strips} of 180 consecutive images at
constant RA  crossing these fields, with $J \leq 13.9$,
and below \emph{or slightly above} the star/galaxy separation ($I/A$) 
threshold
(which depends on the {\sf DENIS} strip)
were selected. The $J$ photometry is based upon the Kron (1980) 
estimator of the
total flux of galaxies. The internal photometric errors, based upon 
multiple observations of single
objects, are $\Delta J \simeq 0.10$ at $J = 13.7$ (Mamon
et al. 1998).
We made use of an early photometric calibration that is
estimated to be accurate to better than $0.2 \,\rm mag$.

Note that our samples do not provide full coverage of the two {\sf UKST}
fields, as only a fraction of {\sf DENIS}
strips  (11 out of 30 for field F787
and 5 out of 27
for field F20) had been observed at the time of the observations.
We ended up with 87 candidate objects in field F787, of which 84 were below
the star/galaxy separation threshold.
Field F20 turned out to be heavily obscured by the Chameleon cloud, and only
33 objects were extracted, all above the star/galaxy separation threshold
(i.e. as galaxies according to {\sf DENIS}).

We did not target objects that were obvious stars, as well as objects that
lied close to others, because of constraints on the minimum fibre separation,
leaving us with 78 targeted objects in field F787 and 32 objects in field
F20.

For all objects, $B$ magnitudes were extracted from the
{\sf ROE/NRL UKST COSMOS Catalog} (hereafter {\sf COSMOS}).
\footnote{See
{\tt
http://xip.nrl.navy.mil/www\_rsearch/RS\_form.html}}
The cross-identification was performed with a $6''$ search radius for
galaxies in astrometrically calibrated {\sf DENIS} strips (strip number 
greater than 5000),
and with a $30''$ search radius otherwise.
In nearly all cases, the  
{\sf DENIS} and {\sf COSMOS} astrometries matched to
within $3''$ (even among the non-calibrated {\sf DENIS} images). 
All cases with no matches within $3''$ were inspected visually
on {\sf DENIS} and {\sf Digitized Sky Survey} (hereafter, {\sf DSS}) images.
In field F787, 6 objects had astrometric matches within
$5''$ to $12''$.
In field F20, one object with no astrometric match turned out to be a false
{\sf DENIS} detection, caused by a nearby very bright and saturated star.
3 objects had {\sf COSMOS} counterparts within $5''$ to $10''$ and one object
had its {\sf COSMOS} counterpart at $69''$.
Comparison with the {\sf DSS} showed that the astrometric offsets were caused
by the lack of calibration of many of the {\sf DENIS} images.
In particular, the mismatch of $69''$ was caused by the poor telescope pointing
caused by a large camera flexure at the end of a polar {\sf DENIS} strip, with
the telescope operating at large zenith angle.

\section{Observations and data reduction}

The spectroscopic observations were performed with the {\sf FLAIR~II}
92-fibre spectrograph at the {\sf UKST} Schmidt telescope at Siding Spring.
The telescope has $1.24\,\rm m$ diameter aperture, and uses $6\farcs7$ fibres.
We made use of the low spectral resolution 250B grating, with a spectral
resolution of $6.14\,\rm \AA / pixel$, observing in the
spectral range $4180\,\rm \AA - 7726\,\rm \AA$.
The spectrograph is described in detail in Parker \& Watson (1995).

The fibres of the plate-holders were configured, off telescope, 
during day-time, with about 11
fibres reserved for sky-background measures to facilitate sky-subtraction.


\begin{table}[ht]
\caption{Journal of FLAIR II observations}
\label{tab_obs}
\bigskip
\centering
\tabcolsep 3.5pt
\begin{tabular}{lccccccc}
\hline
\multicolumn{1}{c}{Date}        & Field  & RA (B1950) & Dec (B1950)& grating & exposure & seeing & weather \\
\hline
1998 March 1  & F787 & 12h00m00s & $-05^\circ00'00''$ & #250B & 17\,400\,s & $2-3''$ & cloud \\
1998 March 4  &  F20 & 12h00m00s & $-80^\circ00'00''$ & #250B & 21\,500\,s & $1-2''$ & cloud \\
1998 March 6  &  F20 & 12h00m00s & $-80^\circ00'00''$ & #250B & #8\,200\,s & $2-3''$ & cloud \\
1998 March 7  &  F20 & 12h00m00s & $-80^\circ00'00''$ & #250B & 14\,500\,s & $2-4''$ & cloud \\
\hline
\end{tabular}
\end{table}

Out of the seven allocated nights, we were only able to observe two fields
with the {\sf FLAIR~II} spectrograph because of 
intermittent cloud cover.
The details of the observations are given in
Table~\ref{tab_obs}. 
For field F787,
we measured the
radial 
velocities on data taken in a batch of $\rm 4200\,s$ and in a longer batch of
$\rm 17\,400\,s$ including the previous one\footnote{Visual inspection of the
spectra suggests that the two batches have signal-to-noise ratios
proportional to the square root of their respective exposure times, to within
10--20\%.}, while for field F20, we only
attempted redshift measurements on a long batch of $44\,200\,\rm s$.
Sky-subtracted, spectrally calibrated spectra were obtained by
reducing the data  as in Ratcliffe et al. (1998),
using the {\tt dofibers} package in {\sf IRAF} (Tody 1993).  
In particular, dome and twilight flats were used.
Spectral calibration was performed with HgCd arcs.
Glitches were removed.
After some experimentation, we combined different exposures by scaling to
their exposure times. 

The redshifts were obtained as follows.
For absorption-featured spectra, redshifts were obtained using the
cross-correlation task {\sf XCSAO} in the {\sf RVSAO} package (Kurtz \& Mink
1998) within {\sf IRAF},
using a series of template spectra from stars, absorption-line and
emission-line galaxies.
We decided to adopt as the
absorption velocity the one associated with the minimum error
from the cross-correlation against the templates. In the great
majority of cases, this coincided also with the maximum $R$ parameter of
Tonry \& Davis (1979).  The redshifts for the emission line objects
were determined using the {\sf EMSAO} task in {\sf RVSAO}.
{\sf EMSAO} finds emission
lines automatically, computes redshifts for each identified line and
combines them into a single radial velocity with error. Spectra
showing both absorption and emission features were generally measured
with the two tasks {\sf XCSAO} and {\sf EMSAO} and the result with the lower
error used. 

\section{Results}

\subsection{Redshifts}
\label{zsec}
We measured velocities successfully for 70 objects in field F787 and 13
objects in field F20.
One object in field F787 at
$v = 500 \, \rm km \, s^{-1}$ looks like an M star from apparent TiO
absorption bands, and one object in field F20
was also a star ($v = -6 \, \rm km \, s^{-1}$).
The results for the 81 spectroscopically
confirmed galaxies 
are presented in Tables~\ref{f787} and \ref{f20} for fields F787 and F20,
respectively.

For the 20 galaxies of field F787 that have spectroscopic redshifts available
in 
the {\sf NED} database, we have
\[
v_{\rm FLAIR} = v_{\rm NED} - 49 \pm 74 \, \rm km \, s^{-1} \ .
\]
Unsurprisingly, 
there are no redshifts in the low galactic latitude field 
F20 present in the {\sf NED} database.
%
%
\begin{table}[htbp]
\caption{Galaxy redshifts in field F787}
\begin{center}
\small
\tabcolsep 2.1pt
\begin{tabular}{cccccccccccc}
\hline
id & RA & Dec & {\scriptsize\sf COSMOS} & $b_J$ & {\scriptsize\sf DENIS} & $J$ & $v$ & $\delta v$ & $R$ & Cont. &
spec.\\
\cline{2-3}
\cline{8-9}
& \multicolumn{2}{c}{(J2000)} & type & & strip & &
\multicolumn{2}{c}{($\rm km \,s^{-1}$)} & & & type \\
\hline
F787--01 & 11h52m17.6s & --07d05m24.0s & G & 17.60 &  4039 & 13.88 & 19604 & 118 & #4.5 & #105 & abs \\ 
F787--02 & 11h52m28.0s & --04d26m28.1s & G & 16.91 &  3708 & 13.29 & 24315 & 117 & #4.4 & #125 & abs \\ 
F787--03 & 11h52m38.0s & --05d12m25.6s & G & 14.78 &  5570 & 11.60 & #5641 & #23 & 26.1 & #310 & em \\ 
F787--04 & 11h52m42.5s & --05d04m36.2s & G & 17.70 &  4062 & 12.74 & 14675 & #36 & #6.9 & #240 & abs \\ 
F787--05 & 11h52m59.7s & --04d25m36.0s & G & 14.14 &  3882 & 12.66 & #1447 & #71 & 10.0 & #170 & em \\ 
F787--06 & 11h53m00.7s & --04d52m45.3s & G & 17.02 &  5436 & 13.40 & 19676 & #74 & #4.6 & ##55 & abs \\ 
F787--07 & 11h53m01.1s & --06d03m54.4s & G & 15.94 &  3905 & 12.57 & #6584 & #45 & #9.1 & #470 & abs \\ 
F787--08 & 11h53m02.4s & --07d37m30.9s & G & 15.33 &  4031 & 13.07 & #7179 & #64 & 13.7 & #180 & em \\ 
F787--09 & 11h53m40.5s & --03d59m46.3s & G & 13.84 &  3882 & 11.90 & #1545 & #73 & 10.8 & #215 & em \\ 
F787--10 & 11h53m44.6s & --05d25m35.9s & G & 15.92 &  4024 & 12.41 & #5779 & #36 & #9.9 & #480 & abs \\ 
F787--11 & 11h53m48.7s & --05d10m04.2s & G & 13.84 &  3905 & 10.84 & #5682 & #78 & #8.8 & #440 & em \\ 
F787--12 & 11h53m59.0s & --02d47m31.5s & G & 17.18 &  4031 & 13.66 & 26885 & 126 & #2.7 & ##80 & abs \\ 
F787--13 & 11h54m00.9s & --06d12m30.7s & G & 17.10 &  3882 & 13.89 & #5493 & #69 & 12.1 & #130 & abs \\ 
F787--14 & 11h54m04.8s & --05d49m59.8s & G & 17.56 &  5570 & 13.77 & 17188 & #71 & #7.4 & #155 & em \\ 
F787--15 & 11h55m00.9s & --06d05m20.9s & G & 15.08 &  5431 & 11.53 & #7741 & #58 & #7.5 & #770 & em \\ 
F787--16 & 11h55m10.2s & --07d50m37.0s & G & 14.37 &  4031 & 11.00 & #5588 & #32 & #9.6 & 1750 & abs \\ 
F787--17 & 11h55m17.9s & --03d43m34.8s & G & 17.67 &  4039 & 13.90 & 19020 & #59 & #3.6 & ##60 & abs \\ 
F787--18 & 11h55m35.0s & --04d48m43.0s & G & 16.18 &  4039 & 13.38 & 18489 & #76 & #3.7 & #160 & em \\ 
F787--19 & 11h56m07.9s & --05d45m41.1s & G & 16.36 &  3708 & 12.83 & 16938 & #33 & #9.4 & #500 & abs \\ 
F787--20 & 11h56m12.2s & --05d28m59.4s & G & 17.15 &  5436 & 13.58 & 24605 & #69 & #6.3 & #190 & abs \\ 
F787--21 & 11h56m13.7s & --03d40m17.0s & S & 13.16 &  3905 & 13.18 & #6395 & #75 & #5.0 & #140 & em \\ 
F787--22 & 11h56m29.1s & --06d24m56.3s & G & 16.41 &  5436 & 12.81 & 14937 & #27 & #5.4 & #210 & abs \\ 
F787--23 & 11h56m43.0s & --07d24m33.8s & G & 16.49 &  5436 & 13.84 & 10654 & #78 & #2.3 & #160 & abs \\ 
F787--24 & 11h56m47.6s & --05d08m06.3s & G & 17.35 &  4031 & 13.88 & 20070 & 109 & #4.3 & ##70 & abs \\ 
F787--25 & 11h56m48.5s & --04d05m40.6s & G & 15.20 &  3708 & 11.18 & #7845 & #30 & 10.6 & #800 & abs \\ 
F787--26 & 11h57m00.1s & --07d13m13.9s & G & 17.20 &  3882 & 13.67 & 11119 & #49 & #9.0 & #300 & abs \\ 
F787--27 & 11h59m14.4s & --04d05m46.2s & G & 16.37 &  4031 & 11.39 & 14128 & #52 & #5.2 & #155 & abs \\ 
F787--28 & 11h59m23.6s & --03d49m30.5s & G & 16.13 &  3905 & 12.78 & #9508 & #62 & #3.3 & #100 & abs \\ 
F787--29 & 11h59m24.5s & --06d44m12.1s & G & 15.85 &  5431 & 12.17 & #8705 & #32 & #7.8 & #500 & abs \\ 
F787--30 & 11h59m35.2s & --06d53m52.6s & G & 17.46 &  3905 & 13.62 & 37954 & #69 & #7.2 & #125 & abs \\ 
F787--31 & 11h59m35.2s & --06d39m44.3s & G & 17.57 &  5415 & 13.92 & 22016 & 104 & #5.3 & #210 & abs \\ 
F787--32 & 11h59m35.6s & --07d13m47.8s & G & 16.08 &  4031 & 13.03 & 15442 & #64 & #9.4 & #380 & em \\ 
F787--33 & 11h59m36.2s & --06d49m35.2s & S & 14.35 &  4039 & 13.83 & 38514 & #25 & #5.7 & #180 & abs \\ 
F787--34 & 11h59m38.3s & --03d22m39.1s & G & 16.67 &  3905 & 13.15 & #5829 & #81 & #6.2 & #170 & abs \\ 
F787--35 & 11h59m38.9s & --03d40m55.8s & G & 15.47 &  3882 & 12.27 & #5856 & #73 & #0.0 & #320 & em \\ 
F787--36 & 11h59m43.9s & --05d46m05.0s & G & 17.70 &  5431 & 13.96 & 18823 & #30 & #6.3 & #120 & abs \\ 
F787--37 & 12h00m18.4s & --07d10m51.2s & G & 15.54 &  4024 & 12.12 & 20572 & #63 & 10.3 & #310 & abs \\ 
F787--38 & 12h00m18.4s & --04d11m36.7s & G & 16.99 &  3905 & 13.92 & 15760 & #82 & #5.3 & ##70 & em \\ 
F787--39 & 12h00m37.8s & --06d51m21.2s & G & 17.71 &  5570 & 13.74 & 38055 & #33 & #7.4 & #260 & abs \\ 
F787--40 & 12h00m41.1s & --05d45m53.0s & G & 16.71 &  4024 & 13.67 & #7843 & #57 & 13.8 & #120 & em \\ 
F787--41 & 12h02m23.0s & --03d18m22.5s & G & 16.76 &  5415 & 13.09 & 20283 & #94 & #3.3 & ##45 & abs \\ 
F787--42 & 12h02m25.6s & --04d18m21.7s & G & 15.32 &  3882 & 12.82 & #5639 & #55 & #9.7 & #115 & em \\ 
\hline
\end{tabular}
\normalsize
\end{center}
\label{f787}
\end{table}

\addtocounter{table}{-1}

\begin{table}[htbp]
\caption{Galaxy redshifts in field F787 (continued)}
\begin{center}
\small
\tabcolsep 2.1pt
\begin{tabular}{cccccccccccc}
\hline
id & RA & Dec & {\scriptsize\sf COSMOS} & $b_J$ & {\scriptsize\sf DENIS} & $J$ & $v$ & $\delta v$ & $R$ & Cont. &
spec.\\
\cline{2-3}
\cline{8-9}
& \multicolumn{2}{c}{(J2000)} & type & & strip & &
\multicolumn{2}{c}{($\rm km \,s^{-1}$)} & & & type \\
\hline
F787--43 & 12h02m45.2s & --03d47m20.8s & G & 17.25 &  5570 & 13.73 & 19355 & #95 & #3.3 & #135 & em \\ 
F787--44 & 12h02m52.7s & --06d48m21.8s & G & 17.87 &  4039 & 13.84 & 38707 & #84 & #4.2 & #120 & abs \\ 
F787--45 & 12h02m58.4s & --06d50m45.5s & G & 17.35 &  3905 & 13.55 & 38524 & #55 & #4.3 & #170 & abs \\ 
F787--46 & 12h03m38.2s & --06d48m46.3s & G & 17.12 &  4024 & 13.73 & 15723 & 129 & #3.6 & ##95 & abs \\ 
F787--47 & 12h03m51.2s & --04d03m11.9s & G & 17.97 &  4039 & 13.78 & 39454 & 117 & #4.7 & ##50 & abs \\ 
F787--48 & 12h03m53.3s & --07d51m39.7s & G & 16.93 &  3708 & 13.25 & 19152 & #43 & #7.6 & #300 & abs \\ 
F787--49 & 12h03m54.7s & --06d32m20.4s & G & 16.29 &  4031 & 13.44 & #1754 & #66 & #9.0 & ##70 & em \\ 
F787--50 & 12h04m16.5s & --07d10m08.5s & G & 18.01 &  3905 & 13.92 & 55087 & #58 & #5.9 & #190 & abs \\ 
F787--51 & 12h04m23.0s & --05d17m47.6s & G & 15.04 &  5436 & 11.65 & #8032 & #83 & 16.0 & #700 & em \\ 
F787--52 & 12h04m23.7s & --07d44m32.6s & G & 17.51 &  3882 & 13.67 & 20908 & 194 & #2.6 & ##65 & abs \\ 
F787--53 & 12h04m58.9s & --05d34m53.0s & G & 16.91 &  4024 & 13.90 & #7928 & #59 & #7.8 & ##90 & em \\ 
F787--54 & 12h05m00.5s & --05d52m53.1s & G & 16.90 &  5570 & 13.54 & 16730 & #83 & #4.5 & #195 & abs \\ 
F787--55 & 12h05m07.1s & --04d04m53.5s & G & 16.57 &  5431 & 12.94 & #9259 & #65 & #7.8 & #180 & abs \\ 
F787--56 & 12h05m07.8s & --04d12m36.1s & G & 16.59 &  5431 & 13.69 & 14067 & #50 & #5.4 & ##90 & em \\ 
F787--57 & 12h05m12.3s & --03d54m14.4s & G & 14.81 &  5570 & 11.41 & #5604 & #64 & #6.5 & #370 & abs \\ 
F787--58 & 12h05m15.0s & --06d51m18.3s & G & 15.63 &  4062 & 13.44 & #7770 & #35 & 10.0 & #215 & em \\ 
F787--59 & 12h05m27.7s & --04d16m05.8s & G & 16.12 &  4031 & 12.63 & #9189 & #82 & 10.4 & #110 & em \\ 
F787--60 & 12h05m25.8s & --04d24m42.5s & S & 14.26 &  5431 & 13.75 & 24227 & #46 & #5.1 & #290 & em \\ 
F787--61 & 12h05m31.9s & --07d38m25.9s & G & 17.08 &  3882 & 13.45 & 20937 & #42 & #5.9 & #205 & abs \\ 
F787--62 & 12h05m33.1s & --04d30m23.0s & G & 16.95 &  5436 & 13.42 & 24428 & 111 & #3.9 & #130 & abs \\ 
F787--63 & 12h05m34.3s & --06d46m18.1s & G & 17.51 &  5415 & 13.96 & 41359 & 137 & #3.1 & #125 & abs \\ 
F787--64 & 12h05m39.0s & --07d05m28.9s & G & 17.78 &  4062 & 13.94 & 37682 & #84 & #6.8 & #220 & abs \\ 
F787--65 & 12h06m37.6s & --06d33m40.3s & G & 16.62 &  3882 & 13.70 & 23139 & #52 & #0.0 & #210 & em \\ 
F787--66 & 12h06m40.6s & --04d26m21.5s & G & 17.45 &  3882 & 13.87 & 25298 & #44 & #8.7 & #100 & abs \\ 
F787--67 & 12h06m43.4s & --04d55m12.2s & G & 17.14 &  5431 & 13.55 & 32250 & 150 & #3.3 & ##55 & abs \\ 
F787--68 & 12h06m55.4s & --07d08m23.6s & G & 16.72 &  4062 & 13.77 & 18670 & #90 & #4.8 & ##95 & em \\ 
F787--69 & 12h07m03.5s & --04d20m51.1s & G & 17.24 &  5570 & 13.45 & 24678 & #81 & #7.8 & #210 & abs \\ 
\hline
\end{tabular}
\end{center}

\noindent Notes: 
The astrometry is taken from the {\sf COSMOS} catalogue.
`{\sf COSMOS} type' is the classification by the {\sf ROE/NRL UKST COSMOS
Catalog}: `G' for galaxy and `S' for star;
$v$ is the measured radial velocity and $\delta v$ is its uncertainty;
$R$ is the Tonry \& Davis (1979) correlation parameter;
`Cont.' gives the maximum continuum, typically reached around $5500\,\rm
\AA$, in arbitrary units;
`spec. type' gives the spectral type: `em' for the presence of
emission lines, otherwise `abs' (for absorption lines only).
Visual inspection of the photographic {\sf UKST}
plates and cross-identification with {\sf NED} indicated the following:\\
\vspace{-1.2\baselineskip}
\begin{center}
\begin{tabular}{ll}
F787--05 = VV 457. & F787--35 has a star superposed on the galaxy. \\
F787--09 = NGC 3952. & F787--37 is within $42''$ of the cluster Abell 1434. \\
F787--14 is a binary galaxy. & F787--49 is a low surface brightness galaxy.\\ 
F787--16 = NGC 3967. & F787--61 is a galaxy triplet.\\
\end{tabular}
\end{center}
\end{table}

\begin{table}[htbp]
\caption{Galaxy redshifts in field F20}
\begin{center}
\small
\tabcolsep 2.5 pt
\begin{tabular}{cccccccccccc}
\hline
id & RA & Dec & {\scriptsize\sf COSMOS} & $b_J$ & {\scriptsize\sf DENIS} & $J$ & $v$ & $\delta v$ & $R$ & Cont. &
spec.\\
\cline{2-3}
\cline{8-9}
& \multicolumn{2}{c}{(J2000)} & type & & strip & &
\multicolumn{2}{c}{($\rm km \,s^{-1}$)} & & & type \\
\hline
F20--01 & 11h20m54.0s & --78d44m30.6s & G & 17.72 & 3674 & 13.21 & #4974 & #35 & 5.1 & 250 & em \\ 
F20--02 & 11h21m06.3s & --78d25m02.1s & G & 16.46 & 3674 & 12.48 & 11508 & #42 & 9.9 & 800 & em \\ 
F20--03 & 11h21m59.4s & --80d21m26.5s & G & 17.52 & 3674 & 13.58 & 11467 & #96 & 3.5 & 170 & abs \\ 
F20--04 & 11h23m50.7s & --79d58m34.8s & G & 18.04 & 3674 & 13.56 & 23328 & #41 & 4.8 & 280 & abs \\ 
F20--05 & 11h25m01.6s & --79d31m35.8s & G & 18.25 & 3674 & 13.70 & 21779 & #67 & 2.8 & 400 & abs \\ 
F20--06 & 11h28m36.3s & --78d17m53.0s & G & 17.84 & 3988 & 13.69 & 18805 & #12 & 9.9 & 480 & em \\ 
F20--07 & 11h29m22.7s & --80d36m30.9s & G & 16.69 & 3988 & 13.43 & 22798 & #92 & 2.7 & 205 & abs \\ 
F20--08 & 11h53m38.8s & --78d56m24.2s & G & 16.80 & 3628 & 13.37 & 12124 & #43 & 9.9 & 500 & em \\ 
F20--09 & 11h53m57.0s & --80d42m07.1s & G & 17.87 & 3628 & 13.42 & 13163 & #91 & 5.3 & 480 & abs \\ 
F20--10 & 12h02m30.9s & --81d15m03.3s & G & 17.83 & 3701 & 13.74 & 13549 & 123 & 3.7 & 180 & abs \\ 
F20--11 & 12h04m02.1s & --78d49m54.2s & G & 17.36 & 3701 & 13.68 & 21229 & #18 & 9.9 & 500 & em \\ 
F20--12 & 12h08m44.4s & --78d45m43.5s & G & 17.97 & 3624 & 12.49 & 29692 & #62 & 6.1 & 570 & abs \\ 
\hline
\end{tabular} 
\normalsize
\end{center}
\vspace{-0.5\baselineskip}
\noindent Notes: 
Columns are the same as in Table 2.
\label{f20}
\end{table}

\subsection{Reliability of the preliminary DENIS galaxy pipeline}
\label{secrelia}
Among the 87 objects extracted by {\sf DENIS} in field F787, 
all but two appeared as 
galaxies upon visual inspection
of the {\sf UKST} photographic plate, while two appeared star-like.
These two stellar objects were classified as stars by {\sf DENIS}.
Note that the third object classified as a star by {\sf DENIS} turned out to
be a galaxy upon visual inspection of the {\sf UKST} plate.
Hence, from visual inspection, the {\sf DENIS} galaxy extraction was
100\% reliable, but incomplete.
Among the 85 visually classified galaxies, 
78 were targeted with {\sf FLAIR}. The resultant spectra produced
69 galaxy redshifts, one stellar spectrum and 8 spectra for which no
redshift could be determined.
In particular, 
all three galaxies that {\sf DENIS} found to be very close to the
star/galaxy separation threshold were targeted and
spectroscopically confirmed as galaxies.

In any event, the reliability of the {\sf DENIS} galaxy extraction for the
high latitude field F787 is between $69/78 = 88\%$ and $68/69 = 99\%$, with a
preference for the higher value, given the visual classifications.

In comparison, 4 of the 84 {\sf DENIS} candidate 
galaxies of field F787, as well as the galaxy (from visual inspection) that
was classified as stellar by {\sf DENIS}
were associated with stars
according to {\sf COSMOS}. Each of these 5 {\sf COSMOS} ``stars''
appeared like galaxies upon visual inspection of the photographic plate, and
subsequent checks with {\sf DENIS} and {\sf DSS} images.
3 of these 5 {\sf COSMOS} ``stars''
were spectroscopically confirmed as galaxies (objects
F787--21, F787--33, F787--60), while
one did not produce a redshift and one was not targeted.
All 5 of these {\sf COSMOS} ``stars'' visually identified as galaxies had
very bright {\sf COSMOS} magnitudes ($b_J < 15$), 
given their $J$ magnitudes, i.e. appeared very
blue. 
Systematic flux overestimates and stellar classifications of bright galaxies
by the {\sf ROE/NRL UKST COSMOS 
Catalog} have been already noted in a systematic comparison with {\sf DENIS}
(Mamon 2000). 

Among the 33 objects extracted by {\sf DENIS} in the low galactic latitude
field F20, inspection of the {\sf UKST} photographic plate 
indicated that 2 were stars,\footnote{The current {\sf DENIS} pipeline would
reject these objects for being too close to one of the edges of their image.}
one was too faint to classify, and one had no counterpart (see end of
Sec.~2).
Finally one object
had poor astrometry (see end of Sec.~2) and turned out to be the
same galaxy as another.
This yields a visually estimated reliability of $28/31 = 90\%$.
All but one of the 30 objects that did not appear as stars nor the ghost
were targeted with {\sf
FLAIR~II}.
Among the 29 spectra, 12 were spectroscopically classified as galaxies, 1
as a star, and 17 produced no redshift.
Thus, the reliability of the {\sf DENIS} galaxy extraction for the
low latitude field F20 is between $12/29 = 41\%$ and $12/13 = 92\%$, again
with a preference for the higher value, based upon visual classification.

In comparison, {\sf COSMOS} classified as stars 3 of the 33 galaxies, among
which the two visually classified stars that were not
targeted with {\sf FLAIR~II}. The remaining {\sf COSMOS} ``star'' produced no
redshift.

Note that the two objects with stellar spectra (one in field F787 and one in
field F20) were both classified as galaxies by
{\sf DENIS}, {\sf COSMOS} and appeared as galaxies by visual inspection of
the {\sf UKST} plate and the {\sf DENIS} image, and we suspect that the
stellar spectra were caused by incorrect fibre configurations.

\subsection{Efficiency of FLAIR II}

For field F787, 
among the nominal 
92 {\sf FLAIR II} fibres, 11 were targeted on the sky, while 3
were broken, leaving a total of 78 object fibres.
For our long batch of $\rm 17\,000\,s$ of {\sf FLAIR~II} observations,
we successfully measured 70 redshifts out of 78 candidates, among which 69
were galaxies. Among the 8 failures,
we suspect that one or two fibres were improperly
configured. 

%
%

%

Table~4
summarises the numbers and fractions of successful
redshift measurements for $\rm 4000\,s$ and $\rm 17\,000\,s$ exposures in the
high latitude F787 field.
At the limiting magnitude of our survey ($J \leq 13.9$), we obtained
redshifts for 2/3 of our galaxy sample for $4\,\rm ks$ exposures and
for 90\% of our sample for $17\,\rm ks$ exposures.

\begin{table}[ht]
\begin{center}
\caption{Redshift success rates for long and short exposures in
field F787}
\vspace{0.5\baselineskip}
\begin{tabular}{cccc|cccc}
\hline
$J^{\rm lim}$ & $N$ & \multicolumn{2}{c|}{$P(\rm success)$} & $b_J^{\rm lim}$
& $N$ & \multicolumn{2}{c}{$P(\rm success)$} \\ 
\cline{3-4}
\cline{7-8}
 & & $17\,\rm ks$ & $4\,\rm ks$ & & & $17\,\rm ks$ & $4\,\rm ks$ \\
\hline
13.0 & 22 & 1.00$\pm0.00$ &  0.86$\pm0.08$ & 16.0 & 21 & 0.95$\pm0.05$ &  0.90$\pm0.07$ \\ 
13.3 & 30 & 0.97$\pm0.03$ &  0.83$\pm0.07$ & 16.5 & 30 & 0.97$\pm0.03$ &  0.87$\pm0.07$ \\ 
13.6 & 42 & 0.95$\pm0.03$ &  0.79$\pm0.07$ & 17.0 & 45 & 0.93$\pm0.04$ &  0.76$\pm0.07$ \\ 
13.9 & 70 & 0.90$\pm0.04$ &  0.67$\pm0.07$ & 17.5 & 62 & 0.90$\pm0.04$ &  0.69$\pm0.07$ \\ 
14.2 & 78 & 0.88$\pm0.04$ &  0.65$\pm0.07$ & 18.0 & 77 & 0.88$\pm0.04$ &  0.65$\pm0.07$ \\ 
\hline
\end{tabular}
\end{center}
\vspace{-0.5\baselineskip}
\noindent Notes: $J^{\rm lim}$ and $b_J^{\rm lim}$ are limiting magnitudes in
the NIR $J$ and optical 
$b_J$ wavebands, respectively. 
$N$ represents the total number of targeted objects within the given limiting
magnitude.
$P(\rm success)$ is the probability
of successful galaxy redshift measurement, i.e. the fraction of targeted
objects that led to non-stellar redshifts. 
The error bars are binomial.
\label{f787rates}
\end{table}

\begin{figure}[ht]
\begin{center}
\centerline{\psfig{file=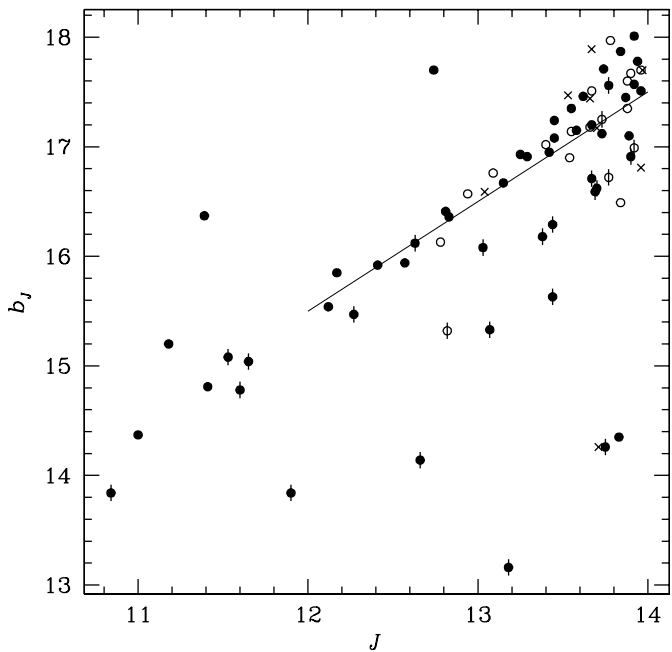,width=10cm}}
\caption{Optical versus near-IR magnitude for galaxies, distinguished by
the success of redshift measurements (field F787).
\emph{Full circles}: redshifts obtained in $\rm 4000\,s$.
\emph{Open circles}: redshifts obtained in $\rm 17\,000\,s$ but not in
$\rm 4000\,s$.
\emph{Crosses}: redshifts not obtained in $\rm 17\,000\,s$.
\emph{Circles with vertical bars}: galaxies with emission lines.
The line indicates a constant colour $b_J - J = 3.5$.
}
\label{velsuccess}
\end{center}
\end{figure}

Figure~\ref{velsuccess} illustrates the success and failure of redshift
measurements in field F787,
using $\rm 4000\,s$ and $\rm 17\,000\,s$ exposures.
The galaxies close to the $J$-band limiting magnitude are much redder than
average, but still allow for redshift measurements.
In fact, we successfully obtained galaxy redshifts in the $17.5 < b_J < 18$
magnitude interval, usually not studied with {\sf FLAIR~II}.
The failures for $17 \rm\, ks$ exposures are not concentrated in
the high $b_J$ region.
This can be caused by either poor $b_J$ photometry, as strongly suggested by
the failure (cross at [13.7,14.2] in Fig.~\ref{velsuccess}) for a galaxy that
is bright in 
$b_J$ but faint in $J$.

Figure~2 provides another look at the success rate of redshift
determinations, by plotting the maximum continuum versus the $J$ magnitude.
\begin{figure}[ht]
\begin{center}
\centerline{\psfig{file=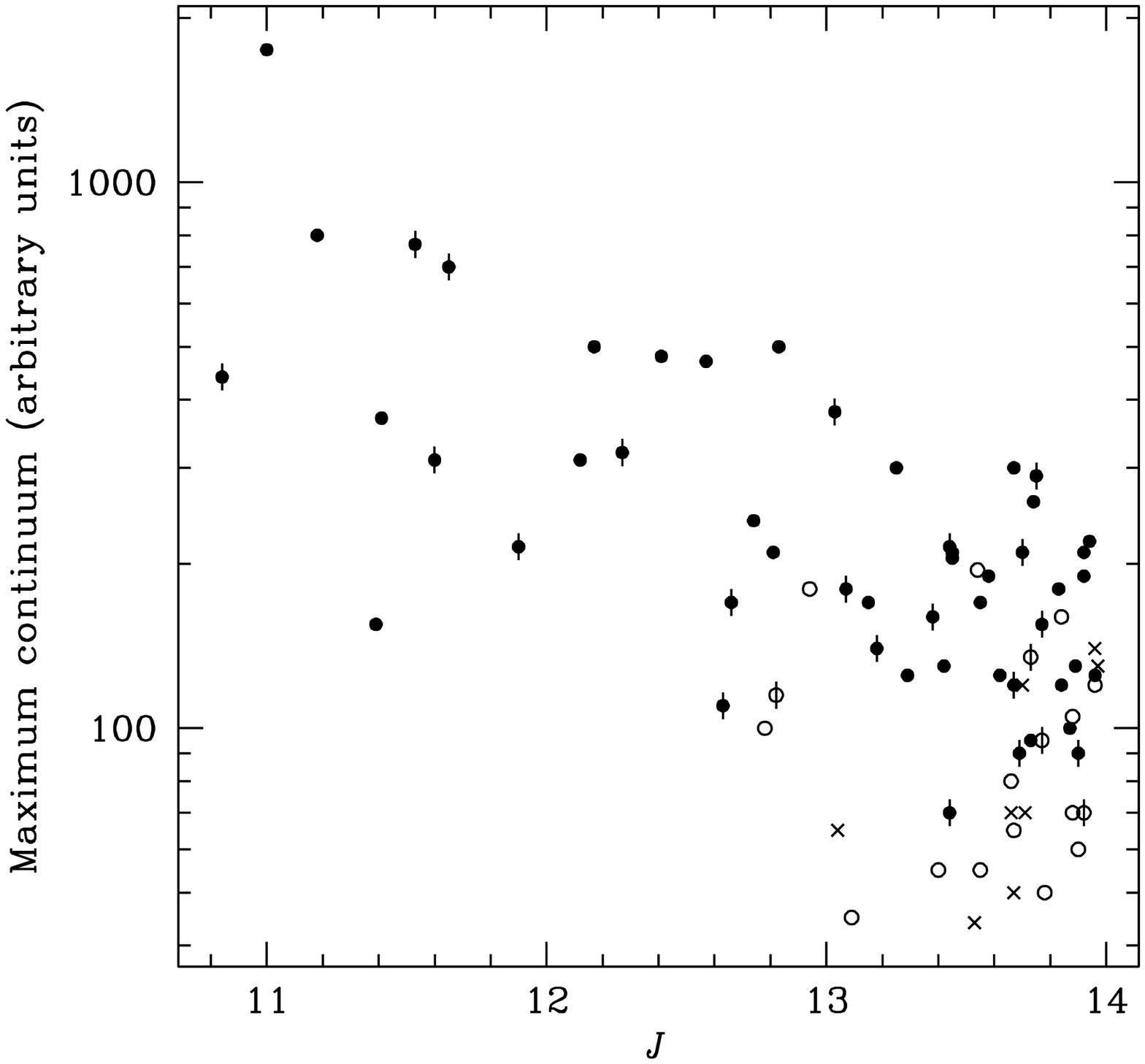,width=10cm}}
\caption{Maximum continuum versus $J$ magnitude for field F787. 
Same symbols as in
Fig.~\ref{velsuccess}.}
\end{center}
\label{contJ}
\end{figure}
The redshift failures are restricted to low continua except for a few
galaxies that are faint in $J$ (and in $B$) with moderate continua.

The large fraction of redshift failures 
may reflect the strong non-uniformity in the fibre
transmissions seen during this run.
Indeed, as illustrated in 
Figure~3,
our dome flats indicated fibre transmissions that differed by factors
of close to 4 in the worse cases.
Among the 8  cases of redshift failure for the
$17\,000\,\rm s$ exposures, 7 were with fibres that had transmissions less
half of the best fibre, and the transmission of the remaining one was lower
than 60\% of that of the best one.
Similarly, for our 18 additional failures for exposures as short as
$4000\,\rm s$ all involved fibres whose transmission was less than 65\% of
that of the best fibre.
In other words, all fibres with transmissions above 65\% of the nominal (best)
value produced 100\% redshift success in field F787.

\begin{figure}[htbp]
\begin{center}
\centerline{\psfig{file=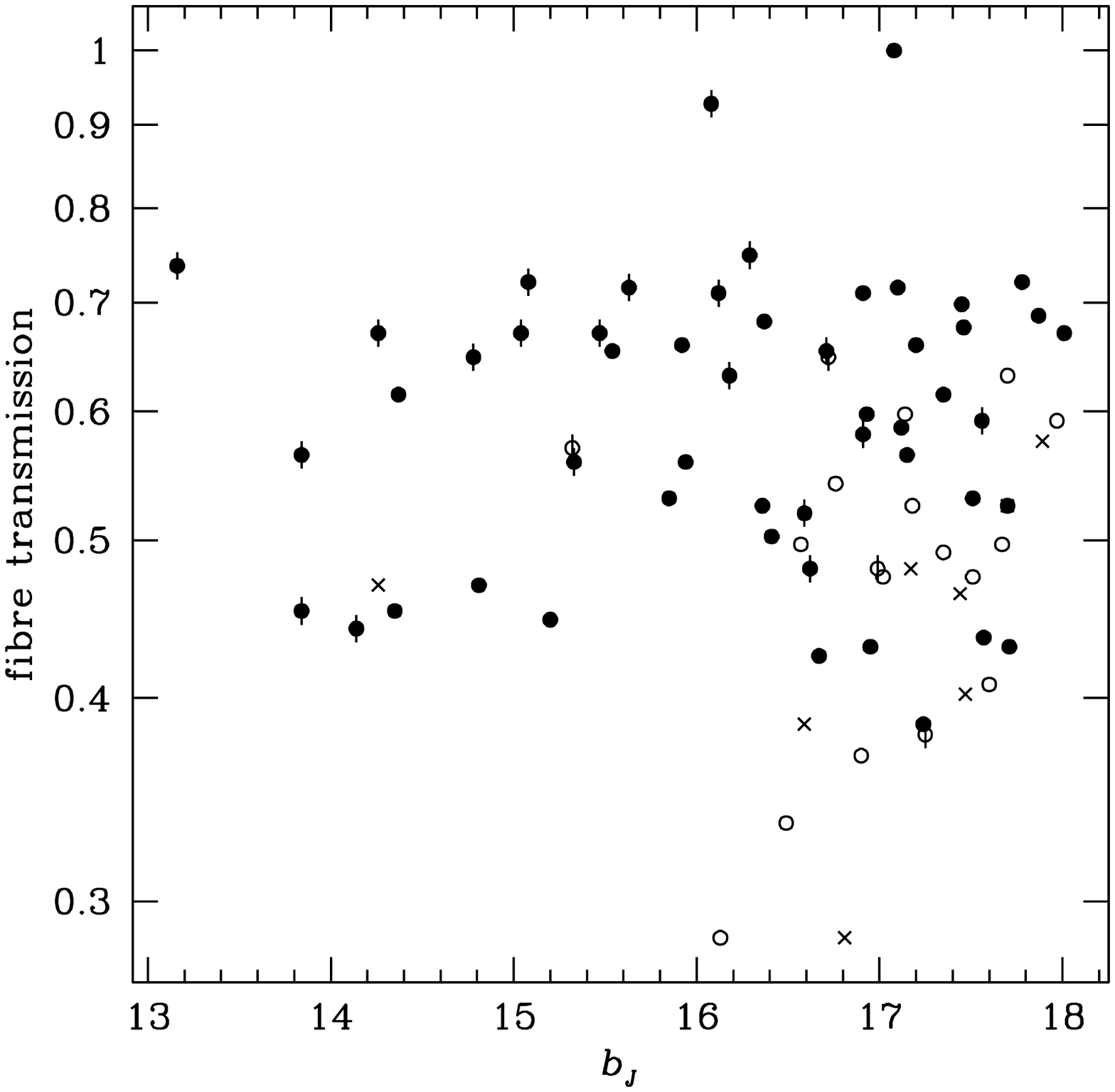,width=10cm}}
\caption{Fibre response (normalised to max=1) versus blue magnitude for field
F787.
Same symbols as in Figure~\ref{velsuccess}.
}
\end{center}
\label{fibresp}
\end{figure}

\begin{figure}[ht]
\begin{center}
\centerline{\psfig{file=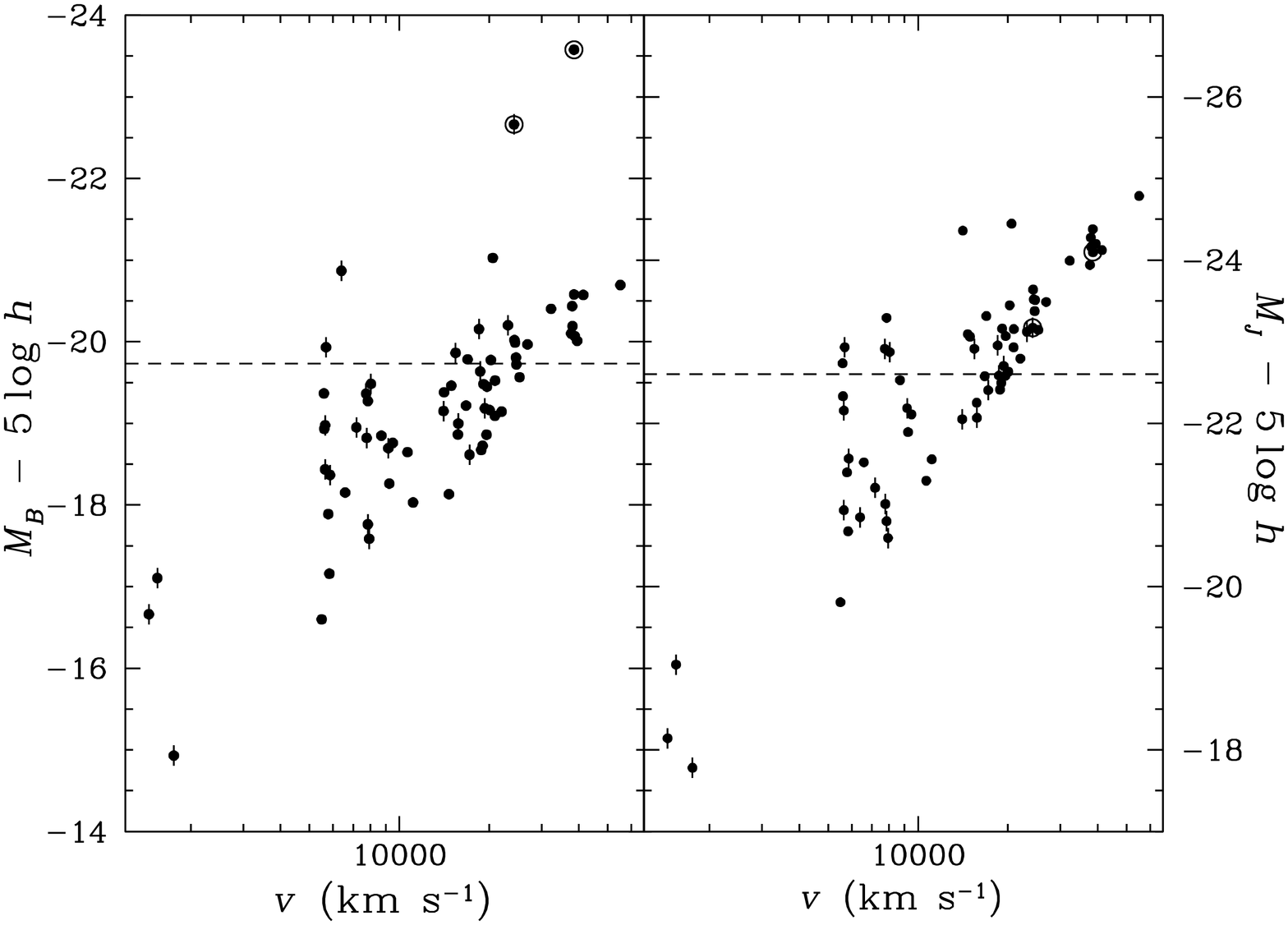,angle=-90,width=15cm}}
\caption{Absolute magnitude versus radial velocity for field F787 
in bands $b_J$
(\emph{left}) and $J$ (\emph{right}).
\emph{Circles with vertical bars}: galaxies with emission lines.
The two galaxies with very bright $B$-band absolute magnitudes are shown as
\emph{surrounded circles}.
The \emph{dashed lines} represent the exponential cutoff of Schechter
functions: $M_B^* = -19.73 + 5\,\log h$ (Folkes et al. 1999) 
and $M_J^* = -22.6 +
5\,\log h$ (eq. [\ref{parsML}] below), where $h = H_0 / (100 \, \rm km \,
s^{-1} \,Mpc^{-1})$.}
\label{vvsBJ}
\end{center}
\end{figure}

Figure~\ref{vvsBJ} shows the absolute magnitudes in $B$ and $J$ versus
radial velocity.
The figure highlights two galaxies with overestimated fluxes in the $B$
band, the more luminous one representing, at face value, a $40\,L_*$ galaxy,
classified as a star in the {\sf COSMOS}, thus
confirming the misclassifications and flux overestimation of bright
galaxies in the {\sf ROE/NRL UKST COSMOS 
Catalog} (see Sec.~\ref{secrelia} above).\footnote{The spectrum of this
object, F787--33, was of good quality, as attested by its large $R$ and small
$\delta v$ in Table~2.}

The rate of overall successful redshift measurement obtained
for field F787 from a $2 \times 2000\,\rm s$ combined integration was a modest
70\% to $b_J<17.5$ increasing to 90\% with a combined field dwell time of
$17\,000\,\rm s$. 
This compares with up to 95\% redshift success reported in the {\sf FLAIR}
manual (Parker 1995; 
Parker \& Watson 1995) for a {\sf COSMOS} $B$ selected sample to
$b_J <17.5$ in $2000\,\rm s$, admittedly obtained immediately after the
system had been overhauled and cleaned.
The
$4000\,\rm s$ integration times are about what is envisaged for the {\sf
6dFGRS} to achieve the
expected survey productivity and completeness. 
Note that the NIR selection did not degrade the rate of redshift success.
Indeed, 
Table~4
indicates a rate of success at $J \leq 13.8 -
13.9$ similar to that of the comparable depth $b_J \leq 17.5$ (the median
$b_J-J$ for our high latitude F787 sample with $J \leq 13.9$ is 3.51).

The efficiency of redshift measurements was much worse for the galaxies in
the low galactic latitude F20 field.
For field F20, among 33 candidate galaxies, only 30 were fibred, while two
objects were bright stars (as seen from inspection of the {\sf UKST} plate)
and one had a detached fibre. 
Among
the 30 spectra, 13 yielded redshifts, one of which turned out to be a star.
As shown in 
Table~5, at all limiting magnitudes,
less than half of the targets yielded
galaxy redshifts for observations as long as $44\,\rm
ks$. This low efficiency is partly due to the increased extinction in the
field.
Indeed, the median $b_J-J$ is 4.22 for the 29 candidate galaxies, in
comparison with a median of 3.51 for the 69 galaxies targeted in field F787,
suggesting $E(b_J-J) = 0.7$, which translates into $A_V \simeq 0.75$.

\begin{table}[ht]
\begin{center}
\caption{Redshift success rates for $44\,\rm ks$ exposures in field F20}
\vspace{0.5\baselineskip}
\begin{tabular}{ccc|ccc}
\hline
$J^{\rm lim}$ & $N$ & $P(\rm success)$ & $b_J^{\rm lim}$ & $N$ & $P(\rm
success)$ \\ 
\hline
13.0 & #5 & 0.40$\pm0.35$ & 17.0 & #7 & 0.43$\pm0.29$ \\ 
13.3 & #8 & 0.38$\pm0.28$ & 17.5 & 10 & 0.40$\pm0.24$ \\ 
13.6 & 16 & 0.50$\pm0.18$ & 18.0 & 24 & 0.42$\pm0.16$ \\ 
13.9 & 29 & 0.41$\pm0.14$ & 18.5 & 27 & 0.44$\pm0.14$ \\ 
\hline
\end{tabular}
\end{center}
\vspace{-0.5\baselineskip}
\noindent Notes: $J^{\rm lim}$ and $b_J^{\rm lim}$ are limiting magnitudes in
the NIR $J$ and optical 
$b_J$ wavebands, respectively. 
$N$ represents the total number of targeted objects within the given limiting
magnitude.
$P(\rm success)$ is the probability
of successful galaxy redshift measurement, i.e. the fraction of targeted
objects that led to non-stellar redshifts. Note that 3 objects had no {\sf
COSMOS} counterparts, hence no optical magnitude.
The error bars are binomial.
\label{f20rates}
\end{table}

Figure~5
illustrates the efficiency of {\sf FLAIR~II} for the
F20 field. The figure indicates that most (but not all) 
failures of redshift measurements
are among red, optically faint ($b_J > 17.5$)
galaxies that suffer from high extinction.
All the counterexamples (except one at $b_J = 17.5$) have transmissions lower
than 55\% of the maximum of the fibres of field F787, so that redshift
failures in field F20 appear to be a combination of extinction and poor fibre
transmission.
\begin{figure}[ht]
\begin{center}
\centerline{\psfig{file=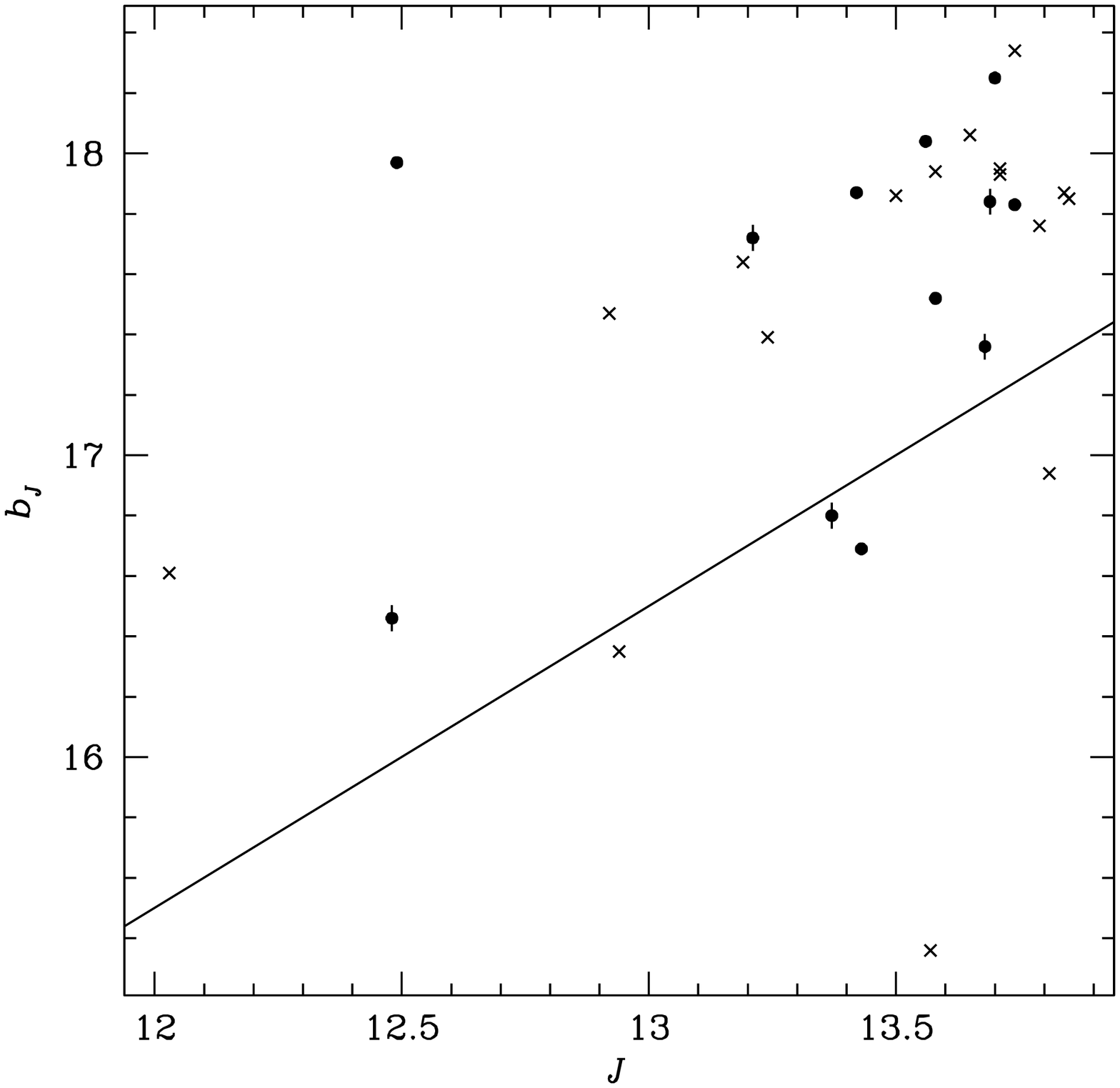,width=10cm}}
\caption{Same as Fig.~1 for the low galactic latitude field F20.
The line represents again $b_J - J = 3.5$ (the median colour at high galactic
latitude) to allow comparison with Fig.~1.}
\end{center}
\label{successf20}
\end{figure}

The lower than expected performance of {\sf FLAIR~II} overall
can be partially explained by
the generally poor seeing and intermittent cloud during the observation nights
adversely affecting the signal-to-noise ratio. 
However, 
Figure~3 strongly suggests that the main culprit was a general decrease in
the fibre 
responses.
The patchy degradation of the overall fibre transmissions and the observed 
wide range in fibre-fibre transmission variations compared to what has been
previously reported for {\sf FLAIR~II} is a known effect associated with the
gradual degradation in the
transmission
of the optical cement used to bond prism to fibre 
caused by the continual high UV illumination of the 
fibre-cement-prism interface when gluing the fibres in position over
the target objects (Lee, 1995, AAO internal report). 
These data were taken at the
end of the refurbishment cycle and shortly after these data were taken,
{\sf FLAIR~II} was
converted to an intermediate magnetic button system as a preliminary
test-bed for {\sf 6dF} and all ferrules
had their prism-cement-fibre interfaces cleaned and
re-done. Significantly
improved fibre-fibre transmission consistency was obtained together
with a recovery in overall fibre transmission.
These improvements bode well for {\sf 6dF}.
Nevertheless, high extinction low galactic latitude fields will require
considerably longer exposure times, and the factor of $44\,200 / 17\,400 =
2.5$ is clearly insufficient.

\subsection{Emission-line galaxies}

An interesting aspect of near-IR selected galaxy samples is that they are not
biased to star forming emission line galaxies (hereafter, ELGs).
It is therefore interesting to know what fraction of a near-IR selected
sample is composed of ELGs.

Figure~\ref{velsuccess} shows that the ELGs in the high galactic latitude
F787 field, represented
with vertical bars, are more frequent in brighter galaxies in the $b_J$
band, but more spread out in $J$ magnitude. The different trend between
optical and NIR bands simply reflects the blue colours of emission-line
galaxies.
This is illustrated in 
Figure~6, which shows that ELGs
are roughly 0.5 magnitude bluer than galaxies that with no emission
lines (hereafter, NELGs) of the same
absolute magnitude.
\begin{figure}[ht]
\begin{center}
\centerline{\psfig{file=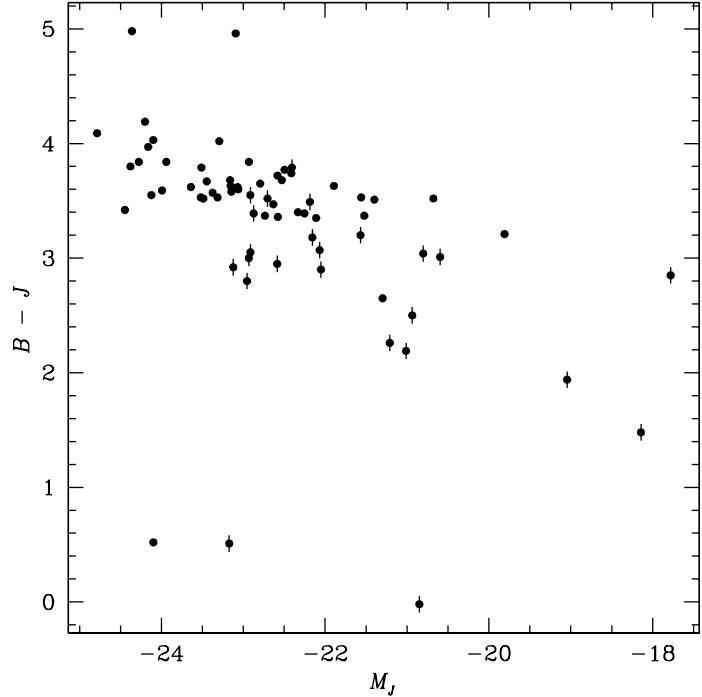,width=10cm}}
\caption{Colour-luminosity diagram for field F787. 
Emission-line galaxies are the 
\emph{symbols
with vertical bars}.
Note that the 3 objects with $b_J - J < 1$ are galaxies classified as stars
by {\sf COSMOS}.
}
\end{center}
\label{colvsMJ}
\end{figure}

Table~6
shows the resulting fractions of ELGs in various
$J$-magnitude limited subsamples of field F787. 
For our largest complete magnitude-limited
sample at $J \leq 13.9$, we have $37\pm6\%$ of emission-line spectra among
our galaxies with measured redshifts.
This fraction can be compared to the 30\% of emission-line spectra in a
spectroscopic followup of {\sf 2MASS} galaxies, selected at $K \leq 12.2$
(Huchra, private communication).
In comparison, over 60\%
of galaxies selected from
two optical samples of comparable depth
(the $b_J < 17.0$ {\sf Durham-UKST} and the
$b_J < 17.5$ {\sf Stromlo-APM} galaxy catalogues)
display H$\alpha$ in emission (Ratcliffe et al. 1998; Tresse
et al. 1999).

\begin{table}[ht]
\begin{center}
\caption{Fraction of emission-line galaxies  
vs. apparent magnitude (field F787)}
\vspace{0.5\baselineskip}
\begin{tabular}{ccc|ccc}
\hline
$J^{\rm lim}$ & $N$ & fraction & $b_J^{\rm lim}$ & $N$ & fraction \\
\hline
12.1 & #9 & 0.56$\pm$0.17 & 15.0 & #9 & 0.67$\pm$0.16 \\ 
12.4 & 12 & 0.50$\pm$0.14 & 15.5 & 15 & 0.73$\pm$0.11 \\ 
12.7 & 16 & 0.50$\pm$0.12 & 16.0 & 20 & 0.60$\pm$0.11 \\ 
13.0 & 22 & 0.41$\pm$0.10 & 16.5 & 29 & 0.55$\pm$0.09 \\ 
13.3 & 29 & 0.41$\pm$0.09 & 17.0 & 42 & 0.52$\pm$0.08 \\ 
13.6 & 40 & 0.38$\pm$0.08 & 17.5 & 56 & 0.41$\pm$0.07 \\ 
13.9 & 63 & 0.37$\pm$0.06 & 18.0 & 68 & 0.35$\pm$0.06 \\ 
\hline
\end{tabular}

\vspace{0.5\baselineskip}
Notes: The error bars are binomial.
\end{center}
\label{emfracvsmag}
\end{table}

In the low galactic latitude
F20 field, one may expect that
selection effects caused by extinction would lead to a different fraction of 
ELGs.
However, the fraction of ELGs in field F20 is $42\pm14\%$,
consistent with that in the high galactic latitude field F787.

Figure~\ref{vvsBJ} shows how the ELGs are distributed in
absolute magnitude and redshift for field F787.
There is a clear trend towards a larger fraction of ELGs for
less luminous galaxies.
This is quantified in 
Table~7.
\begin{table}[ht]
\begin{center}
\caption{Fraction of emission-line galaxies 
vs. absolute $J$ magnitude (field F787)}
\vspace{0.5\baselineskip}
\tabcolsep 0.3cm
\begin{tabular}{ccc}
\hline
$M_J$ & $N$ & $f$ \\
\hline
--25 to --24 &  10 & 0.00$\pm$0.00 \\ 
--24 to --23 &  18 & 0.11$\pm$0.07 \\ 
--23 to --22 &  24 & 0.50$\pm$0.10 \\ 
--22 to --17 &  17 & 0.59$\pm$0.12 \\ 
\hline
\end{tabular}

\vspace{0.5\baselineskip}
Notes: The error bars are binomial.
\end{center}
\label{emfracvsMJ}
\end{table}

Figure~7
shows the distribution of absolute $J$ magnitudes in field F787.
A Wilcoxon rank-sum test indicates over 97\% probability that the
$J$-band luminosities of ELGs are, on average, lower than those of NELGs.
\begin{figure}[ht]
\begin{center}
\centerline{\psfig{file=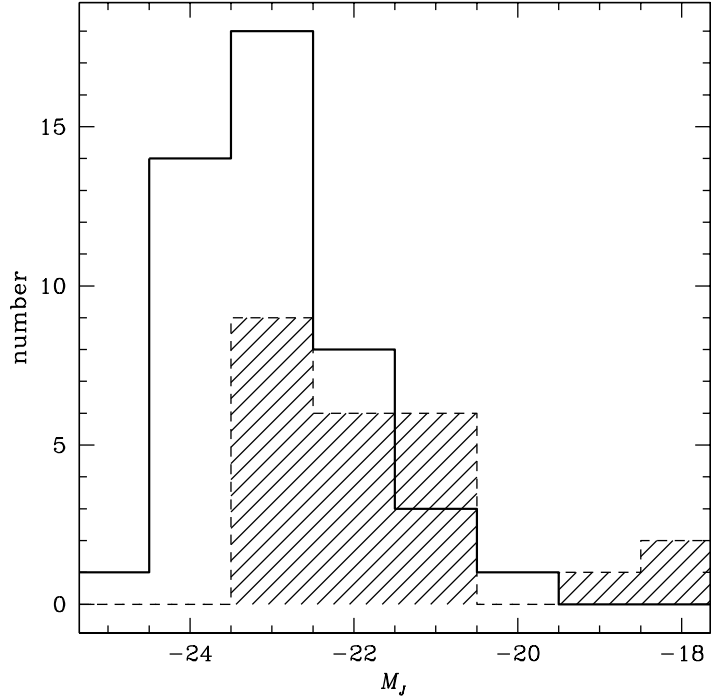,width=10cm}}
\caption{Distribution of absolute $J$ magnitudes for galaxies in field F787
with
(\emph{shaded dashed histogram}) and without 
(\emph{solid histogram}) emission lines.}
\end{center}
\label{nvsMJ}
\end{figure}
A qualitatively similar trend of increasing fraction of ELGs with decreasing
luminosity has been reported by Tresse et
al. (1999) for the optically selected {\sf Stromlo-APM}
sample.
Compared to the optical, one expects a stronger trend of ELG fraction versus
absolute magnitude in the near-infrared, because ELGs are
intrinsically blue (see Fig. 6).

\subsection{Luminosity function}
\label{lfsec}

We have attempted to evaluate the $J$-band galaxy luminosity function for the
63 galaxies with $J \leq 13.9$ of
the high galactic latitude field F787, using the simplest methods given in
Willmer (1997).
Because of the small number of galaxies, we restrain ourselves to the
STY
(Sandage, Tammann \& Yahil 1979) parametric, maximum likelihood estimator,
assuming a Schechter (1976)
luminosity function 
%
$\phi(L) = {\phi_* / L_*}\, (L/ L_*)^\alpha\,\exp(-L/ L_*)$. 

Neglecting
magnitude and redshift errors, as well as galactic extinction and
the $k$ terms (indeed, $|k_J| \leq 0.1$ for $z \leq 0.2$ as in our sample, 
see Poggianti 1997; Mannucci et al. 2001) 
and assuming $\Omega_m=0.3$ and
$\Omega_\Lambda=0.7$ (for the luminosity distances required to convert
apparent to absolute magnitudes),
we derive reasonable
maximum likelihood parameters: 
\begin{equation}
\alpha = -1.1\pm 0.4
 \qquad
M_J^* = -22.6\pm 0.5 + 5\,\log h
\ ,
\label{parsML}
\end{equation}
where the errors are 68\% confidence constant likelihood contours equal to
the maximum likelihood value offset by the usual $0.5\,f_{\rm \chi^2}
(2)$.\footnote{C.~Willmer kindly checked for us our luminosity function, and
furthermore ran Monte-Carlo simulations of magnitude-limited galaxy
distributions with 63 members, 
derived from a parent Schechter luminosity function to check that the
uncertainty in the STY estimate of the parameters was of the order of the
error bars quoted in equation~(\ref{parsML}).}
%
%

We evaluated the normalisation of the luminosity function 
using the relation  (e.g. Willmer 1997, sec. 3) $\phi_* = \bar n /
\int_{M_{\rm bright}}^{M_{\rm faint}} 
\phi(M)\,dM$,
where $\bar n$ is the number density of galaxies (corrected for the selection
function, see Willmer 1997, eq. [33]), evaluated in 10 velocity bins
out to $35\,000 \, \rm km \, s^{-1}$ (beyond which our data is too sparse).
We obtain
\begin{equation}
\phi_* = 0.010_{-0.0035}^{+0.0052}\,h^3\,\rm {Mpc}^{-3} \ ,
\label{phistar}
\end{equation}
where the uncertainty is estimated from the uncertainty of $\bar n$ (thus
neglecting the uncertainties in $\alpha$ and $M_J^*$).
%

Our luminosity function parameters (eqs.~[\ref{parsML}] and [\ref{phistar}]) 
match well those of Cole et al. (2001),
and our slope $\alpha$ and normalisation $\phi_*$ also 
match very well that of the
optical Stromlo-APM survey  (Loveday et al. 1992).

\subsection{Large-scale structure}

\begin{figure}[htbp]
\begin{center}
\centerline{\psfig{file=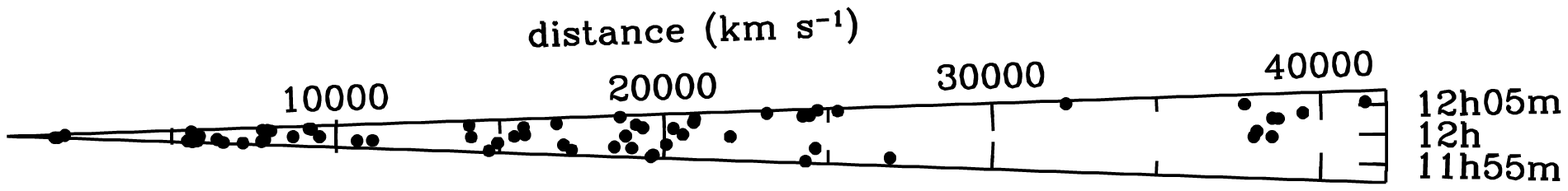,width=15cm}}
\centerline{\psfig{file=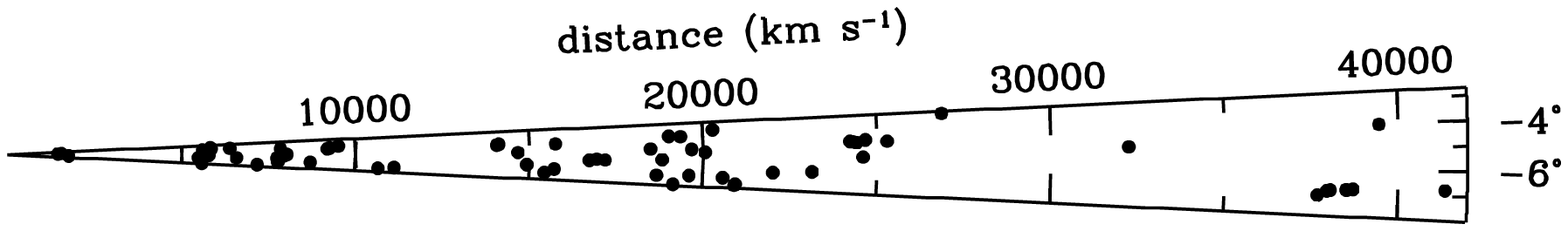,width=15cm}}
\caption{Wedge diagrams
for field F787. One galaxy lies at $v = 55,000 \, \rm
km \, s^{-1}$.}
\label{wedgef787}
\end{center}
\end{figure}

Figure~\ref{wedgef787} shows the resulting wedge
diagrams for field F787. Note that these wedges are incomplete, as only 11
out of 35 possible {\sf DENIS} strips covered the {\sf UKST} F787
field. 
One notices galaxy concentrations  at radial velocities near 20\,000 and
$38\,000 \, \rm km \, s^{-1}$, the first of which includes a galaxy
that lies within $42''$ of the Abell 1434
cluster, whose redshift is not yet known.

Our measurement based upon a single, possibly
interloping object gives a tentative $z = 0.0686$ for Abell 1434.
If Abell 1434 lies at that distance, an angle of $42''$ corresponds to a
physical separation of $42 \, h^{-1} \, \rm kpc$, which would place
object F787--37 at the core of the cluster.

Inspection of Figure~\ref{wedgef787} indicates that
the concentration at $38\,000 \, \rm km \, s^{-1}$ is for a fairly constant
Dec and a range of RAs, corresponding to a comoving scale of $11 \, h^{-1} \,
\rm Mpc$, indicative of a supercluster.

\begin{figure}[htb]
\begin{center}
\centerline{\psfig{file=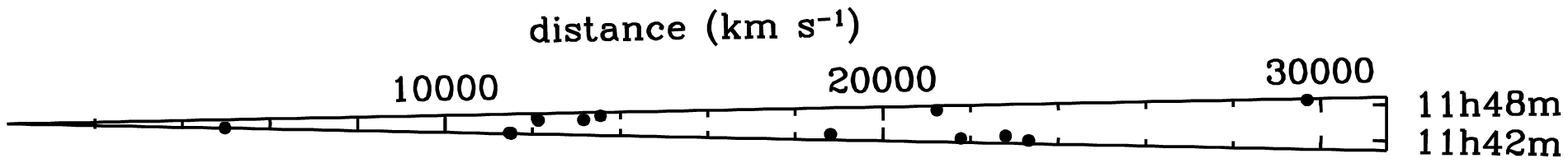,width=15cm}}
\centerline{\psfig{file=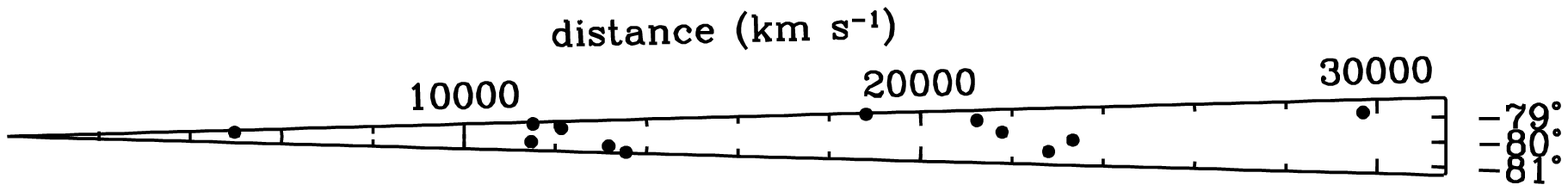,width=15cm}}
\caption{Wedge diagrams for field F20. These wedges are narrower than in
field F787, because no {\sf DENIS} strips were available near the plate edges.}
\label{wedgef20}
\end{center}
\end{figure}

Figure~\ref{wedgef20} show the resulting wedge
diagrams for field F20. Concentrations are apparent at $cz = 12\,000$ and
$22\,000 \, \rm km \, s^{-1}$.

\section{Conclusions}

This pilot study for the forthcoming {\sf 6dFGRS}
has provided useful information on various aspects of the survey.
First, it has shown that the preliminary galaxy extraction from {\sf DENIS}
images was highly reliable, even at low galactic latitudes ($b = -17^\circ$),
where the reliability, based upon subsequent visual inspection of
photographic plates was
93\%, despite confusion with double stars and
extinction.
Second, it pointed to a degraded fibre transmission, which has been attended to
since. Also a newly commissioned 1K EEV CCD for {\sf 6dF} 
is expected to significantly improve
the instrument throughput.
Third, it proved that optical spectroscopy of high-latitude, NIR-based, 
samples is feasible
with similar rates of successful redshift measurement as in
optically-selected galaxy samples.
Fourth, it indicated that redshifts can be obtained at low galactic latitude,
although with much less success than at high galactic latitude.

There are therefore good reasons to have full confidence in the success of
the {\sf 6dFGRS} in obtaining a highly complete redshift
followup of galaxies selected in the NIR by {\sf 2MASS} and {\sf DENIS},
although this success may be limited to high galactic latitudes.
Note that the primary sample of {\sf 6dFGRS} will be the longer
wavelength $K_s$ band at $2.15\,\mu$m, and there is expected to be  a larger
fraction of galaxies with only absorption lines, for
which redshifts are generally harder to measure.

Moreover, our study has provided a first estimate on the fraction of
emission-line galaxies in a $J$-band selected galaxy sample
and its variation with galaxy $J$-band luminosity, as well as an
estimate of the $J$-band galaxy luminosity function.
Serendipitous results include a tentative redshift for the Abell 1434
cluster as well as the establishment of various
concentrations of galaxies, including a potential supercluster at $380 \,
h^{-1} \, \rm Mpc$. 

\section*{Acknowledgements}

We warmly thank Malcolm Hartley for operating the {\sf UKST}
telescope, Paul Cass for
fibering one of the plates, and Matthew Colless and 
Brian Boyle for useful discussions and
encouragement.
We are very grateful to Chris Willmer who checked our luminosity functions
and ran Monte-Carlo simulations to check the error estimates.
We also thank Santiago Arribas and a second anonymous referee for useful
comments on the manuscript.
These observations were based upon images taken
by the \nobreak 
{\sf DENIS} near-infrared imaging survey, which is supported, in
France by the Institut National des Sciences de l'Univers, the Education
Ministry and the Centre National de la Recherche Scientifique, in Germany by
the State of Baden-W\"urtemberg, in Spain by the DG1CYT, in Italy by the
Consiglio Nazionale delle Ricerche, in Austria by the Fonds zur F\"orderung
der Wissenschaftlichen Forschung und Bundesministerium f\"ur Wissenshaft und
Forschung.
We are grateful to the {\sf
DENIS} staff, and in particular Jean Borsenberger for flat-fielding the {\sf
DENIS} images, and Emmanuel Bertin for help with the {\sf SExtractor} source
extraction package.  
GAM acknowledges a travel grant from the {\sf Coop\'eration France-Australie}.
We made use of the {\sf ROE/NRL COSMOS UKST Southern Sky
Object Catalog},
the  {\sf Digitized Sky Survey}, produced at the Space Telescope Science
Institute under US     Government grant NAG W-2166,
 and the {\sf NASA/IPAC Extragalactic Database (NED)}, which is
operated by the Jet Propulsion Laboratory, California Institute of
Technology, under contract with the National Aeronautics and Space
Administration.


\section*{References}







\reference Bertin, E. \& Arnouts, S. 1996, A\&AS 117, 393
\reference Cole, S., Norberg, P., Baugh, C. M., Frenk, C. S. et al. 2001,
MNRAS in press, astro-ph/0012429
\reference Davis, M. \& Huchra, J. P. 1982, ApJ 274, 437
\reference Epchtein, N., Deul, E., Derriere, S. et al. 1999, A\&A 349, 236
\reference Folkes, S., Ronen, S., Price, I. et al. 1999, MNRAS 308, 459
\reference H\'eraudeau, P., Simien, F. \& Mamon, G. A. 1996, A\&AS 117, 417
\reference Jarrett, T. H., Chester, T., Cutri, R. et al. 2000a, AJ 119, 2498
\reference Jarrett, T. H., Chester, T., Cutri, R. et al. 2000b, AJ 120, 298
\reference Kron, R. G. 1980, ApJS 43, 305
\reference Kurtz, M. J. \& Mink, D.J. 1998, PASP 110, 934
\reference Loveday, J., Peterson, B. A., Efstathiou, G. \& Maddox,
S. J. 1992, ApJ 390, 338
\reference Mamon, G. A. 1998, in Wide Field Surveys in Cosmology,,
ed. Colombi, S., Mellier, Y. \& Raban, B., 323, astro-ph/9809376
\reference Mamon, G. A. 2000, in Cosmic Flows 1999: Towards an Understanding
of Large-Scale Structure, ed. Courteau, S., Strauss, M. A. \& Willick, J. A.,
103, astro-ph/9908163
\reference Mamon, G. A., Borsenberger, J., Tricottet, M. \& Banchet, V. 
1998, in The Impact of Near-Infrared Sky Surveys on
Galactic and Extragalactic Astronomy, ed. Epchtein, N., Kluwer,
177, astro-ph/9712169 
\reference Mannucci, F., Basile, F., Poggianti, B. M., Cimatti, A., Daddi, E.,
Pozzetti, L. \& Vanzi, L., 2001, MNRAS, submitted, astro-ph/0104427
\reference Parker, Q. A. 1995, Spectrum 7, 17
\reference Parker, Q. A. \& Watson F. G. 1995, in Wide Field Spectroscopy and
the Distant Universe, ed. Maddox, S. J., World Scientific, 33
\reference Parker, Q. A., Watson, F. G. \& Miziarski, S. 1998, in
Fiber-Optics in Astronomy III, ASP Conference Series
Volume 152, ed. Arribas, S., Mediavilla, E. \& Watson, F., 80
\reference Poggianti, B. M. 1997, A\&AS 122, 399
\reference Ratcliffe, A., Shanks, T., Parker, Q. A., Broadbent, A., Watson,
F. G., Oates, A. P., Collins, C. A. \& Fong, R., 1998, MNRAS 300, 417
\reference Sandage, A., Tammann, G. A. \& Yahil. A. 1979, ApJ 232, 352
\reference Schechter, P. 1976, ApJ 203, 297
\reference Tody, D. 1986, SPIE 627, 733
\reference Tonry J. \& Davis, M. 1979, AJ 84, 1511 
\reference Tresse, L., Maddox, S. J., Loveday, J. \& Singleton, C. 1999,
MNRAS 310, 262
\reference Watson, F. G., Parker, Q. A. et al. 2000, SPIE 4008, 123
\reference Willmer, C. N. A. 1997, AJ 114, 898
\reference Zaritsky, D., Rix, H.-W. \& Rieke, M. 1993, Nat 364, 313


\end{document}